\documentstyle[12pt]{article}
\def\be{\begin{eqnarray}}
\def\ee{\end{eqnarray}}
\def\ba{\begin{array}}
\def\ea{\end{array}}
\def\p{\phi}
\def\vp{\varphi}

\def\eps{\epsilon}

\def\pa{\partial}

\def\n{\nabla}

\def\N{{\cal N}}
\def\G{{\cal G}}
\def\B{{\cal B}}

\def\X{{\cal X}}
\def\H{{\cal H}}

\def\5{^{(5)}}
\begin{document}
\begin{center}
{\LARGE {The Inverse Scattering Method, Lie--B\"acklund\\
\vskip 0.2cm Transformations and Solitons for Low--energy\\
\vskip 0.25cm Effective Field Equations of 5D String Theory}}
\end{center}
\vskip 1.5cm
\begin{center}
{\bf \large {Alfredo Herrera--Aguilar}}
\end{center}
\begin{center}
Theoretical Physics Department, Aristotle University of Thessaloniki\\
54006 Thessaloniki, Greece\\
and Instituto de F\'\i sica y Matem\'aticas, UMSNH\\
Edificio C--3, Ciudad Universitaria, Morelia, Mich. CP 58040 M\'exico\\
e-mail: aherrera@auth.gr
\end{center}
\vskip 0.1cm
\begin{center}
and
\end{center}
\vskip 0.1cm
\begin{center}
{\bf \large {Refugio Rigel Mora--Luna}}
\end{center}
\begin{center}
Escuela de Ciencias F\'\i sico--Matem\'aticas, UMSNH\\
Edificio B, Ciudad Universitaria, Morelia, Mich. CP 58040 M\'exico\\
e-mail: rigel@fismat.umich.mx
\end{center}

\vskip 1.5cm
\begin{abstract}

In the framework of the 5D low--energy effective field theory of
the heterotic string with no vector fields excited, we combine two
non--linear methods in order to construct a solitonic field
configuration. We first apply the inverse scattering method on a
trivial vacuum solution and obtain an stationary axisymmetric
two--soliton configuration consisting of a massless gravitational
field coupled to a non--trivial chargeless dilaton and to an axion
field endowed with charge. The implementation of this method was
done following a scheme previously proposed by Yurova. We also
show that within this scheme, is not possible to get massive
gravitational solitons at all. We then apply a non--linear
Lie--B\"acklund matrix transformation of Ehlers type on this
massless solution and get a massive rotating axisymmetric
gravitational soliton coupled to axion and dilaton fields endowed
with charges. We study as well some physical properties of the
constructed massless and massive solitons and discuss on the
effect of the generalized solution generating technique on the
seed solution and its further generalizations.
\end{abstract}
\newpage
\section{Introduction}

The aim of this paper is to apply two non--linear methods for
constructing solitonic solutions in the framework of the 5D
truncation of the low--energy effective field theory of the
heterotic string with no vector fields excited. In a first stage
we implement the inverse scattering method (ISM) on a trivial
(flat space--time) seed solution in order to obtain a solitonic
solution. At this step we make use of a chiral $SL(4,{\bf
R})/SO(4)$ representation of the stationary axisymmetric theory
and construct a rotating massless gravitational soliton following
the scheme proposed by Yurova in \cite{yurova} for chiral matrices
of dimension greater than two. Afterwards we endow this object
with gravitational mass and dilaton charge by means of a
non--linear transformation of Lie--B\"acklund type, the so--called
normalized Ehlers transformation (NET).

In the framework of General Relativity, Belinski and Zakharov
\cite{bz} demonstrated that the vacuum stationary axially
symmetric gravitational equations written in chiral form may be
integrated with the aid of the ISM. This technique allows one to
obtain the $N$--soliton configuration starting from flat
space--time by making use of a symmetric chiral $2\times2$--matrix
which must not satisfy any group condition. It was shown, in
particular, that the Kerr--NUT metric can be interpreted as a
two--soliton solution of 4D Einstein theory in the presence of two
commuting Killing vectors. As far as we know, the ISM has not been
generalized for gravitational systems involving three space--time
variables, i.e. with just one Killing vector imposed. Thus, in
order to apply the ISM to gravitational models and their
extensions we must consider configurations that depend at most on
two space--time coordinates.

In the same way, in the framework of D--dimensional low--energy
effective string theories toroidally reduced down to two
space--time dimensions, i.e. in the presence of $D-2$ Killing
vectors, we hope to obtain black hole (or black brane) solitonic
solutions by applying the ISM. However, in this realm the problem
becomes more complicated due to the fact that, in general, the
chiral representations of the reduced low--energy effective field
theories have dimensions greater than two and must satisfy,
indeed, non--trivial group conditions. For instance, the chiral
model which describes the low--energy effective field theory of
the heterotic string when reduced to three space--time dimensions
possesses the $SO(d+1,d+n+1)/[SO(d+1)\times SO(d+n+1)]$ symmetry
group \cite{ms}, where $d$ is the number of compactified
dimensions and $n$ is the number of Abelian vector fields of the
theory\footnote{In fact, the further reduction of the theory to
two space--time dimensions by imposing one more Killing vector
does not increase the dimensionality of the chiral
representation.}. Thus, the chiral representation of this theory
involves symmetric matrices of dimension $(2d+n+2)$ which must
satisfy orthogonal group conditions. As a consequence, the
original scheme described in \cite{bz} cannot be applied anymore
and must be suitably modified.

In fact, it is not so easy to overcome this difficulty since both,
the dimensionality of the chiral matrices and their group symmetry
condition strongly restrict the solitons we can obtain, leading
sometimes to rotating massless gravitational configurations
\cite{yurova} and, even more, to trivial solutions (see below).
For example, in the framework of the scheme followed in this
paper, this problem reduces to the choice of five relationships
between eight constant parameters. However, not every set of such
conditions leads to a soliton configuration depending on three
parameters. Moreover, among them there are conditions that lead to
trivial objects, i.e., to flat space--time solutions. Thus, this
work brings some insight into the understanding of obtaining such
solitonic objects, always with the hope of clarifying all the
conditions under which we can apply the ISM and construct the
general $N$--soliton solution for systems represented by chiral
matrices of any dimension greater than two which, indeed, must
satisfy non--trivial symmetry group conditions (see the related
works \cite{bakyur}, where the ISM has been implemented for
special systems in the framework of string theory).

Another interesting issue concerns the physical interpretation of
the constructed soliton since it describes a massless
gravitational object\footnote{An object without mass term in the
asymptotical expansion of the $g_{tt}$--component of the metric
tensor.} coupled to non--trivial dilaton and Kalb--Ramond fields.
However, despite the massless character of the gravitational
configuration, it possesses angular momentum, a strange feature
that must be clarified. In this context, it is also interesting to
see whether or not it is possible to endow our two--solitonic
configuration with gravitational mass and dilaton charge since
there is no way to get a massive solution when implementing the
ISM on a trivial solution within the scheme proposed by Yurova
(see Sec. III). In order to achieve this goal we apply a solution
generating technique based on the use of a non--linear
transformation of Lie--B\"acklund type \cite{stephani}, namely, we
perform the NET \cite{hk5} on our massless two--soliton and get a
rotating gravitational configuration with mass term coupled to
dilaton and axion fields endowed with their respective charges.

It is with this motivation that we perform the present
investigation. The paper is organized as follows: In Sec. II we
present the 5D low--energy effective action of the theory under
consideration as well as the matrix Ernst potential (MEP)
formulation of the reduced down to three dimensions theory. Then
we recall the NET of the stationary theory in the language of the
MEP and write down an alternative $SL(4,{\bf R})/SO(4)$
representation of the stationary theory. In Sec. III we describe
the ISM and the modifications that one must perform in order to
apply it to our string system. Afterwards we show that, within the
scheme proposed by Yurova, one can construct just massless
gravitational solitons and present an explicit exact solution. We
also analyze some physical properties of this two--soliton object.
In  Sec. IV we perform a simplified NET on a seed solution which
corresponds to the solitonic configuration constructed in Sec. III
and get a gravitational soliton endowed with mass term. We study
as well some physical properties and limits of the obtained
solution. Finally, in Sec. V we summarize our results and discuss
on the physical peculiarities of the constructed solutions and
also analyze the technical details of the implemented non--linear
methods that produce them. Here we give as well some suggestions
concerning the further development and generalization of the
techniques applied in this work.

\section{Low--energy Effective Action}
We shall study the 5D low--energy effective field theory of the
heterotic string with no vector fields excited. This theory is
described by the following action \be S\5\!=\!\int\!d^5\!x\!\mid\!
G\5\!\mid^{\frac{1}{2}}\!e^{-\p\5}\left(R\5\!+\!
\p\5_{;M}\,\p^{(5);M}
\!-\!\frac{1}{12}\!H\5_{MNP}\,H^{(5)MNP}\right), \ee where
$H\5_{MNP}=\pa_MB\5_{NP} + \mbox{{\rm cycl. perms. of}
M,\,N,\,P;}$ $G\5_{MN}$ is the metric, $B\5_{MN}$ is the
anti--symmetric Kalb--Ramond field, $\p\5$ is the dilaton and
$M,N,P=1,2,...,5$. After the Kaluza--Klein compactification of
this model on $T^2$ \cite{ms}, the resulting stationary theory
possesses the $SO(3,3)/[SO(3)\times SO(3)]$ symmetry group and
describes the three--dimensional dilaton field $\p$ and the scalar
$2\times2$--matrices $G\equiv G_{pq}$ and $B\equiv B_{pq}$ \be
\p=\p\5-\frac{1}{2}{\rm ln}\mid{\rm det}G\mid, \qquad
G_{pq}=G_{p+3,q+3}\5, \qquad B_{pq}=B_{p+3,q+3}\5, \label{GBp} \ee
where $p,q=1,2$ label the time and extra coordinates,
respectively; the vector fields represented by the $2\times
3$--matrices $(A_1)_{\mu}^p$ and $(A_2)_{\mu}^{p+2}$ \be
(A_1)_{\mu}^p=\frac{1}{2}\left(G^{-1}\right)_{pq}G_{q+3,\mu}\5,
\qquad (A_2)_{\mu}^{p+2}=\frac{1}{2}B_{p+3,\mu}\5-B_{pq}A_{\mu}^q,
\ee where $\mu,\nu=1,2,3$ are the dynamical coordinates; and the
antisymmetric tensor field $B_{\mu\nu}$ (which we set to zero from
now on due to its nondynamical properties in three dimensions) \be
B_{\mu\nu}=B_{\mu\nu}\5-4B_{pq}A_{\mu}^pA_{\nu}^q-
2\left(A_{\mu}^pA_{\nu}^{p+2}-A_{\nu}^pA_{\mu}^{p+2}\right); \ee
all these fields are effectively coupled to three--dimensional
gravity which is described by the metric tensor \be
g_{\mu\nu}=e^{-2\p}\left[G_{\mu,\nu}\5-
G_{p+3,\mu}\5G_{q+3,\nu}\5\left(G^{-1}\right)_{pq}\right]. \ee

It turns out that for stationary configurations, the vector fields
can be dualized on--shell through the pseudoscalar fields $u$ and
$v$ as follows
\begin{eqnarray}
\nabla\times\overrightarrow{A_1}&=&\frac{1}{2}e^{2\p}G^{-1}
\left(\nabla u+B\nabla v\right),
\nonumber                          \\
\nabla\times\overrightarrow{A_2}&=&\frac{1}{2}e^{2\p}G\nabla v-
B\nabla\times\overrightarrow{A_1},\label{dual}
\end{eqnarray}
where all vector and differential operations are performed with
respect to the metric $g_{\mu\nu}$.

Thus, the effective stationary theory describes gravity
$g_{\mu\nu}$ coupled to the scalars $G$, $B$, $\p$ and the
pseudoscalars $u$, $v$.

\subsection{Matrix Ernst Potentials}
All these matter fields can be arranged in the following matrix
\be \X=\left( \ba{cc} -e^{-2\p}+v^TXv&v^TX-u^T \cr Xv+u&X \ea
\right), \label{X}\ee where $X=G+B$. This is a $3\times3$--matrix
which was called matrix Ernst potential in \cite{hk3} because of
the close analogy existing between the representation the
low--energy effective field theory of the heterotic string and the
formulation of the stationary Einstein--Maxwell (EM) theory in
terms of the complex Ernst potentials \cite{e}. Its components
have the following physical meaning: the relevant information
about the gravitational field is contained in the matrix potential
$X$ through the matrix $G$, whereas its rotational character is
encoded in the dualized variable $u$; $X$ also parameterizes the
antisymmetric Kalb--Ramond tensor field $B$, whereas its
multidimensional components are dualized through $v$, finally, the
3D dilaton is $\phi$. In terms of the MEP the effective stationary
theory adopts the form \be ^3S\!= \!\int\!d^3x\!\mid
g\mid^{\frac{1}{2}}\!\{\!-\!R\!+ \!{\rm
Tr}[\frac{1}{4}\left(\nabla \X\ \right)\!\G^{-1} \!\left(\nabla
\X^T\right)\!\G^{-1}]\},\label{acX}\ee where
$\G=\frac{1}{2}\left(\X+\X^T\right)$ is the symmetric part of the
matrix potential $\X$, whereas the antisymmetric one reads
$\B=\frac{1}{2}\left(\X-\X^T\right)$. The MEP $\X$ can be
expressed as the sum of its symmetric and antisymmetric parts:
$\X=\G+\B$ \, where \be \G=\left( \ba{cc} -e^{-2\p}+v^TGv&v^TG \cr
Gv&G \ea \right), \qquad \B=\left( \ba{cc} 0&v^TB-u^T \cr Bv+u&B
\ea \right). \label{GyB}\ee

\subsection{The Normalized Ehlers Transformation}

In the language of the MEP the stationary action (\ref{acX})
possesses a set of symmetries which has been classified according
to their charging properties in \cite{hk5}. Among them we find the
so--called normalized Ehlers and Harrison transformations, NET and
NHT respectively, which act in a non--trivial way on the
space--time when applying solution generating techniques. For
instance, in the framework of General Relativity, the Ehlers
transformation generates the NUT parameter when applied to both
Schwarzschild and Kerr solutions, whereas the Harrison
transformation endows these metrics with electromagnetic charges.
We would like to mention that the NET constitutes a matrix
generalization of the charging symmetry of Lie--B\"acklund type
introduced by Ehlers in the framework of General Relativity
\cite{kramer}.

The matrix NET transformation reads \cite{hk5} \be
\X\rightarrow\left(1+\Sigma\Lambda\right)
\left(1+\X_0\Lambda\right)^{-1} \X_0\left(1-\Lambda\Sigma\right)+
\Sigma\Lambda\Sigma,\label{net} \ee where $\Sigma=diag(-1,-1;1)$
and $\Lambda$ is an arbitrary antisymmetric constant $3\times
3$--matrix parameter $\Lambda=-\Lambda^T$. In the framework of
solution generating techniques, by applying the NET on a
stationary seed solution we obtain a new stationary solution
endowed with three more parameters introduced through the
antisymmetric matrix $\Lambda$. We shall apply this technique in
Sec. IV on a solitonic seed solution constructed by means of the
ISM in order to further analyze the physical effect of the NET and
the physical properties of the generated solution.

\subsection{$SL(4,{\bf R})/SO(4)$ Chiral Representation of the Model}
In \cite{ky1} it was pointed out that apart of
$SO(3,3)/[SO(3)\times SO(3)]$ symmetry group formulation, the
stationary system under consideration allows an alternative
$SL(4,{\bf R})/SO(4)$ chiral parametrization. The latter
formulation is more convenient since it is parameterized by
$4\times4$--matrices instead of $6\times6$--matrices, and has to
satisfy a trivial group condition. Thus, the chiral action reads
\be ^3S\!= \!\int\!d^3x\!\mid g\mid^{\frac{1}{2}}\!
[\!-\!R\!+\!\frac{1}{4}\!{\rm
Tr}\left(J^{\N}\right)^{2}],\label{acN}\ee where
$J^{\N}=\n\N\N^{-1}$, the symmetric matrix $\N$ is \be
\N=\left({\rm det}\G\right)^{-1/2}\left( \ba{cc} \G& \G\H\cr
\H^T\G&{\rm det}\G+\H^T\G\H \ea \right), \label{N}\ee the
$3\times1$--column $\H$ is determined by the relation
$\H_k\eps_{mnk}=\B_{mn}\equiv\B$, the matrix potentials $\G$ and
$\B$ are defined in (\ref{GyB}), \,$\eps_{mnk}$ is the
antisymmetric tensor with $\eps_{123}=1$ and $m,n,k=1,2,3$. The
matrix $\N$ is indeed unimodular and belongs to the $SL(4,{\bf
R})/SO(4)$ group.

Let us consider an axially symmetric field configuration. In this
case the spatial metric can be written in the Lewis--Papapetrou
form \be
ds_3^2=g_{\mu\nu}dx^{\mu}dx^{\nu}=e^{2\gamma}(d\rho^2+dz^2)+\rho^2d\vp^2,
\ee the effective action of the system can be expressed as follows
\be ^2S= \frac{1}{4}\int d\rho dz\rho{\rm
Tr}\left(J^{\N}\right)^{2},\label{ac2N}\ee and the matrix field
equation reads \be \n\left(\rho J^{\N}\right)=0.\label{meqs} \ee
The corresponding Einstein equations determine the function
$\gamma$ through the relations \be \gamma_{,z}=\frac{\rho}{4}{\rm
Tr}\left(J^{\N}_{\rho}J^{\N}_z\right), \qquad
\gamma_{,\rho}=\frac{\rho}{8}{\rm
Tr}\left[\left(J^{\N}_{\rho}\right)^2-\left(J^{\N}_z\right)^2\right],
\ee where the operator $\n$ is related to the flat two--metric
$\delta_{ab}$ and all dynamical variables depend on $\rho$ and $z$
only. These equations are automatically satisfied once a solution
for (\ref{N}), or equivalently (\ref{GyB}), is found. In the next
Sec. we shall construct a solitonic solution to this matrix
equation by means of the ISM.

\section{Solitons Via ISM}
In this Sec. we shall continue to apply the inverse scattering
technique in order to construct soliton solutions for the 5D
low--energy effective field theory under consideration. However,
because of the symmetry condition that the real chiral matrix $\N$
must satisfy, the original scheme of Belinski and Zakharov is not
applicable anymore and the implementation of such a powerful
method turns out to be not so simple. Another problem is related
to the dimension of the chiral matrix for which solitons are
constructed. The point here is that the matrix $\N$ must possess
certain asymptotic properties which correspond to concrete
physical values that the fields must adopt at spatial infinity.
Thus, the bigger the dimension of the matrix $\N$ is, the more
boundary conditions we must impose on it. In this context we shall
avoid these difficulties following the modification of the ISM
proposed by Yurova. Thus, the integration of the field equations
(\ref{meqs}) is associated with the L--A pair \be
D_1\psi=\frac{\rho J^N_z-\lambda J^N_\rho}{\lambda^2+\rho^2}\psi,
\qquad  D_2\psi= \frac{\rho J^N_\rho+\lambda
J^N_z}{\lambda^2+\rho^2}\psi,\label{LA} \ee where $J^N=\rho
J^{\N}$ and the differential operators are \be D_1=\partial_{z}-
\frac{2\lambda^{2}}{\lambda^{2}+\rho^{2}}
\partial_{\lambda}, \qquad  D_2=\partial_{\rho}+
\frac{2\lambda\rho}{\lambda^{2}+\rho^{2}}
\partial_{\lambda}, \label{d1d2}
\ee $\lambda$ is a spectral complex parameter and $\psi$ =
$\psi(\lambda, \rho, z)$. The solution of (\ref{meqs}) for the
symmetric matrix $\N$ constitutes the function $\psi$ with
vanishing value of the spectral parameter, i.e., \be \N(\rho, z) =
\psi (0, \rho, z). \ee Thus, for any known solution $\psi_0$ of
the system (\ref{LA})--(\ref{d1d2}), the function $\psi$ can be
obtained in the form \be \psi=\chi\psi_0, \label{psi} \ee where
the equations for $\chi$ are \be D_1\chi=\frac{\rho J^N_z-\lambda
J^N_\rho}{\lambda^2+\rho^2}\chi-\chi\frac{\rho (J^N_z)_0-\lambda
(J^N_\rho)_0}{\lambda^2+\rho^2}, \qquad  D_2\chi=\frac{\rho
J^N_\rho+\lambda J^N_z}{\lambda^2+\rho^2}\chi-\chi\frac{\rho
(J^N_\rho)_0+\lambda (J^N_z)_0}{\lambda^2+\rho^2}. \nonumber\ee

The matrix $\N$ must be real and symmetric; in order to ensure its
real character, we shall impose the condition \be
\chi(\lambda)=\bar\chi(\bar\lambda) \label{real}\ee (see \cite{bz}
for details). However, in order to apply the ISM to chiral
matrices with dimension greater than two, the symmetry requirement
must be imposed after the construction of the solitonic solution.
Thus, after the implementation of ISM, the constructed matrix $\N$
will not be symmetric and we must impose this condition
afterwards.

The soliton solutions for the matrix $\N$ correspond to the pole
divergences in the spectral parameter complex plane for the
matrices $\chi$ and $\chi^{-1}$. When the poles are simple, these
matrices can be represented as follows: \be
\chi=I+\sum_{k=1}^{N}\frac{R_k}{\lambda-\mu_k},  \qquad
\chi^{-1}=I+\sum_{k=1}^{N}\frac{S_k}{\lambda-\nu_k},
\label{chichi} \ee where the pole trajectories for each pole $k$
are determined by \be \mu_k(\rho,z)= w_{(\mu)}-z \pm
[(w_{(\mu)}-z)^{2}+{\rho}^{2}]^{\frac{1}{2}}, \qquad w_{(\mu)}=
{\rm const.} \label{poles} \ee for $\mu_k(\rho,z)$ and the same
equation for $\nu_k(\rho,z)$ with constants $w_{(\nu)}$. From
$\chi\chi^{-1}=I$ in the poles $\mu_{k}$ and $\nu_{k}$ it follows
that \be R_k\chi^{-1}(\mu_k)=S_k\chi(\nu_k)=0. \label{RS} \ee
Hence the matrices $R_k$ and $S_k$ are degenerate and can be
represented as follows \be (R_k)_{ab}=n^{k}_{a}m^{k}_{b}, \qquad
(S_k)_{ab}=p^{k}_{a}q^{k}_{b}. \label{nmpq} \ee By substituting
Eqs. (\ref{chichi}) and (\ref{nmpq}) into Eq. (\ref{RS}) we obtain
\be n^{k}_{a}=\sum^{N}_{l=1}p^{l}_{a}{{\Gamma}_{kl}}^{-1}, \qquad
q^{k}_{a}=-\sum^{N}_{l=1}m^{l}_{a}{{\Gamma}_{kl}}^{-1}, \quad
\mbox{{\rm where}} \quad
{{\Gamma}_{kl}}=\frac{\sum_{c}p^{k}_{c}m^{l}_{c}}{\mu_{l}-\nu_{k}},
\label{nq} \ee

\be m^{k}_{a}=[\psi^{-1}_{0}(\mu_{k},\rho,z)]_{ca}m^{k}_{c0},
\qquad \mbox{{\rm and}} \qquad
p^{k}_{a}=[\psi_{0}(\nu_{k},\rho,z)]_{ac}p^{k}_{c0}, \label{mp}\ee
where $m^{k}_{c0}$ and $p^{k}_{c0}$\, are arbitrary constants and
$a,b,c=1,2,3,4$.

When considering the two--soliton case, in \cite{yurova} it was
shown that in order to have a unimodular matrix $\N$, the relation
\be \mu_1\mu_2=\nu_1\nu_2 \label{mu1mu2}\ee is really important
and it constitutes the main difference with respect to the scheme
proposed by Belinski and Zakharov since it is not compatible with
the symmetry requirement
$\N=\chi(-\rho^2/\lambda)\N_0\chi^T(\lambda)$ of \cite{bz}. As a
consequence, the resulting unimodular matrix $\N$ will be
non--symmetric. However, the symmetry conditions may be attained
by a suitable choice of the arbitrary parameters of (\ref{mp}).

Let us apply the modified ISM in order to construct a stationary
axially symmetric two--soliton solution for the 5D string model
which is described by the effective action (\ref{ac2N}) when
reduced to two dimensions.

In the simplest case, the initial values of the metric and field
variables correspond to flat space--time. Thus, the seed chiral
matrix adopts the form \be \N_0=\mbox{\rm
diag}(-1,-1;1,1).\label{N0}\ee We shall construct the solitonic
solution using the set of coordinates of Boyer and Lindquist
without mass. This is due to the fact that the resulting
two--soliton gravitational solution corresponds to a massless
source (see bellow for details). Thus, we have \be
\rho=\left(r^{2}-\sigma^{2}\right)^{\frac{1}{2}}\sin{\theta},
\qquad z-z_{1}=r\cos{\theta}, \ee where the new constants
$\sigma=\frac{1}{2}(w_{(\mu)}-w_{(\nu)})$ and
$z_{1}=\frac{1}{2}(w_{(\mu)}+w_{(\nu)})$. Consequently, the pole
trajectories read \be \mu_1=2\sin^2\frac{\theta}{2}(r+\sigma),
\qquad \mu_2=-2\cos^2\frac{\theta}{2}(r-\sigma),\nonumber\ee \be
\nu_1=-2\cos^2\frac{\theta}{2}(r+\sigma), \qquad
\nu_2=2\sin^2\frac{\theta}{2}(r-\sigma), \label{4poles}\ee and
obviously satisfy the condition (\ref{mu1mu2}). Since
$\N_0=\psi_0^{-1}(\mu_k,\rho,z)=\psi_0(\nu_k,\rho,z)$, then the
vectors $p_a^k$ and $m_a^k$ constitute arbitrary constants (see
Eq. (\ref{mp})). Thus, by applying the scheme described above, one
can construct the matrix two--soliton solution $\N$ for Eq.
(\ref{meqs}) depending on these vectors. Such a matrix will be
unimodular, but not symmetric. The following conditions provide
the symmetric character of the matrix $\N$: \be m_a^k=p_a^k,
\label{restrics1}\ee for all $a,k$; and \be p_3^2=p_3^1, \qquad
p_2^2=\left(p_3^1\right)^2/p_2^1, \qquad
p_4^2=p_4^1=-p_1^1p_3^1/p_2^1, \qquad
p_1^2=\left(p_3^1/p_2^1\right)^2p_1^1. \label{restrics2}\ee Thus,
only three constants survive the symmetrization of the matrix
$\N$. However, we would like to point out that just the condition
(\ref{restrics1}) is not enough to ensure the symmetry character
of $\N$, it leaves eight arbitrary constants, but it does not lead
to a symmetric matrix $\N$. Moreover, we must impose five more
restrictions on these eight constants in order to obtain the
desired symmetric matrix $\N$. The choice of these restrictions is
not unique and the constants are not independent each other.
Moreover, some choices lead to a trivial soliton solution or to a
solution that depends on just two arbitrary constants instead of
three. In this respect, we observe some differences between our
choice of constants and the one performed by Yurova. For example,
by setting our $p_1^1$ to zero we just set to zero one parameter
of the solution, whereas within the choice made in \cite{yurova},
the author claims that if any of the constants $p_a^k$ is set to
zero, all other constants vanish as well and the constructed
solution turns out to be a trivial one. This fact, in turn, leads
to slightly different solitonic solutions from the physical point
of view (see below).

\subsection{Massless Character of the Gravitational Soliton}

Now we shall show that when constructing the two--soliton solution
and imposing the condition (\ref{restrics1}) towards the
symmetrization of the matrix $\N$, we necessarily obtain a
massless gravitational field configuration. Thus, after taking
into account the restrictions (\ref{restrics1}), we obtain a
chiral matrix $\N$ which has the following block structure and
asymptotic behaviour \be \N_{as}\equiv \left( \ba{cc} N_1&N_2 \cr
N_3 &N_4 \ea \right)=\left( \ba{cccc}
-1&n_{12}/r&n_{13}/r&n_{14}/r \cr n_{21}/r&-1&n_{23}/r&n_{24}/r\cr
n_{31}/r&n_{32}/r&1&n_{34}/r \cr n_{41}/r&n_{42}/r&n_{43}/r& 1 \ea
\right), \label{Nas} \ee where $N_1$ is a $3\times3$--matrix,
$N_2$ is a $3\times1$--column, $N_3$ is a $1\times3$--row and
$N_4=1$, \be
n_{ac}=2\kappa\sigma\frac{p_a^1p_c^2-p_c^1p_a^2}{p_b^1p_b^2r},
\quad a\ne c, \quad \mbox{and} \quad \kappa=\left\{ \ba{rcc} -1,&
&a=1,2. \cr 1, & &a=3,4. \ea \right. \ee On the other side, the
matrix $\N$ is defined through $\G$ and $\B$ through the relation
(\ref{N}). By equating both representations we can, in principle,
compute the {\it non--symmetric} matrix
$\G_{as}=\left(N_4-N_3N_1^{-1}N_2\right)N_1$ which asymptotically
behaves as follows \be \G_{as}=\left( \ba{ccc}
-1&n_{12}/r&n_{13}/r \cr n_{21}/r&-1&n_{23}/r \cr
n_{31}/r&n_{32}/r&1 \ea \right) \label{Gas} \ee and obviously does
not possess Coulomb terms in the diagonal components. Thus, when
looking for constructing a massive gravitational solitonic
configuration within this concrete modified version of the ISM, it
does not matter what kind of relationships we choose between the
remaining eight constants $p_a^k$, since the conditions
(\ref{restrics1}) are sufficient to restrict us to obtain massless
solitons.

Thus, by imposing the conditions (\ref{restrics1}), it is not
possible at all to obtain a massive gravitational two--soliton
solution. This fact also shows how strong, from the physical point
of view, are the restrictions we must impose on our matrix $\N$ in
order to make possible the application of the ISM to chiral
matrices of dimension greater than two. It remains an open
question whether a different scheme of symmetrization of the
matrix $\N$ can lead to massive gravitational solitons in the
framework of the considered model. A current research in this
direction is being performed by the authors.

\subsection{Explicit Two--soliton Solution}

Once the matrix $\N$ is constructed taking into account the
conditions (\ref{restrics1}) and (\ref{restrics2}), we can come
back to the original variables of the theory with the aid of
(\ref{N}), (\ref{GyB}) and (\ref{GBp})--(\ref{dual}). At this
stage it is convenient to introduce the following notations \be
a=\frac{\sigma\left[\left(p_3^1\right)^2+\left(p_2^1\right)^2\right]}{2p_2^1p_3^1},
\qquad
b=\frac{\sigma\left(p_1^1\right)^2\left[\left(p_3^1\right)^2-\left(p_2^1\right)^2\right]}
{2p_2^1p_3^1\left[\left(p_1^1\right)^2+\left(p_2^1\right)^2\right]},
\qquad c=\frac{\sigma
p_2^1\left[\left(p_3^1\right)^2-\left(p_2^1\right)^2\right]}
{2p_3^1\left[\left(p_1^1\right)^2+\left(p_2^1\right)^2\right]},
\label{abc}\ee which establishes the following relation
$\sigma^2=a^2-(b+c)^2$; let us define as well \be
\Delta=r^2-\sigma^2, \qquad \delta^2=r^2+c^2-(b-a\cos\theta)^2.
\ee

Now we shall write down the final expression of the constructed
two--soliton by implementation of the ISM. The 5D line element
reads \be
ds_5^2=G_{pq}(dx^p-\omega_{\vp}^pd\vp)(dx^q-\omega_{\vp}^qd\vp)+e^{2\p}ds_3^2,
\label{interval}\ee where the components of the matrix $G_{pq}$
have the form \be
G_{11}=-\frac{r^2+b^2-(c-a\cos\theta)^2}{\delta^2}, \quad\quad
G_{12}=\frac{2cr}{\delta^2}, \quad\quad
G_{22}=\frac{r^2+b^2-(c+a\cos\theta)^2}{\delta^2}, \label{Gpq}\ee
the metric functions $\omega_{\vp}^q$ are \be
\omega_{\vp}^1=\frac{-2a\sqrt{bc}\,r\sin^2\theta}{\Delta+a^2\sin^2\theta},
\quad\quad \omega_{\vp}^2=2\sqrt{bc}\left[\cos\theta+
\frac{a\left(b+c-a\cos\theta\right)\sin^2\theta}{\Delta+a^2\sin^2\theta}\right],
\label{oms}\ee the three--dimensional dilaton field is \be
e^{2\p}=1-\frac{4bc}{\Delta+a^2\sin^2\theta}, \label{3dilaton}\ee
and the expression for the spatial line element is the following
\begin{eqnarray}
ds_3^2=\left(\Delta+a^2\sin^2\theta\right)\left[\frac{dr^2}{\Delta}+
d\theta^2\right]+\Delta\sin^2\theta d\varphi^2.
\end{eqnarray}
The components of the antisymmetric Kalb--Ramond tensor field are
defined by the relations \be
B_{12}\!=\!\frac{2br}{\delta^2},\quad\,
B_{4,\vp}\5\!=\!\frac{2\sqrt{bc}\,r(2b\cos\!\theta\!+\!a\sin^2\!\theta)}{\delta^2},\quad\,
B_{5,\vp}\5\!=\!2\sqrt{bc}\!\left[\!\cos\!\theta\!-
\!\frac{a(c\!-\!b\!+\!a\cos\!\theta)\!\sin^2\!\theta}{\delta^2}\!\right]\!\!.
\label{B12}\ee Finally, the 5D dilaton field reads \be
e^{\p\5}=\frac{r^2+(b-c)^2-a^2\cos^2\theta}{\delta^2}.
\label{5dilaton}\ee

These expressions describe an stationary axially symmetric
massless gravitational field configuration coupled to a
non--trivial dilaton field without charge and to an axion field
endowed with charge. A novel feature of this configuration is that
it possesses angular momentum. By analyzing the asymptotical
behaviour of the functions $\omega_{\vp}^q$ we observe that
$\omega_{\vp}^1$ defines the angular momentum according to the
following relation \be \omega_{\vp}^1\mid_{r\rightarrow\infty}
\,\,\,\sim\frac{-2a\sqrt{bc}\sin^2\theta}{r}, \label{om1}\ee
whereas $\omega_{\vp}^2$ is not an asymptotically flat quantity
since at spatial infinity it behaves as \be
\omega_{\vp}^2\mid_{r\rightarrow\infty} \,\,\,\sim
2\sqrt{bc}\cos\theta \label{om2}\ee and defines a NUT--like
parameter. However, it is worth noticing that in order to obtain
an asymptotically flat field configuration we can set to zero
either the parameter $b$ or $c$. In the first case the remaining
configuration constitutes a massless static inhomogeneous
gravitational field with non--trivial components of the matrix
$G_{pq}$ with no dilaton and Kalb--Ramond fields excited. In the
second case the truncated configuration is different, it
represents a massless static inhomogeneous gravitational field
with non--zero components $G_{11}$ and $G_{22}$ coupled to a
non--trivial massless 5D dilaton field and endowed with an axion
field which possesses a charge term. One can notice another quite
strange property of our soliton: as soon as we require asymptotic
flatness, the solution becomes static (see Eqs. (\ref{om1}) and
(\ref{om2})). In this respect, the soliton constructed in
\cite{yurova} has quite different properties since in order to get
an asymptotically flat field configuration we must set to zero
both parameters $b$ and $c$, obtaining a trivial solution in this
way. Thus, that solitonic solution does not contain asymptotically
flat field configurations.

In both of the considered limits ($b=0$ and $c=0$) we obtained
static inhomogeneous field configurations even when the parameter
$a$, which usually is the responsible for the rotation of the
gravitational field, is not vanishing. Thus, our solitonic
configuration does not contain a spherically symmetric subclass of
solutions (in accordance with the soliton constructed in
\cite{yurova}). Comparison to other results in the literature
shows \cite{othersols} that this solutions are not obtained by
setting to zero the mass or other parameters. Of course, it is
interesting to study other physical properties of these massless
solitonic configurations. At first glance it seems that these
solutions correspond to both rotating or static inhomogeneous
black strings since we have the presence of horizons, however,
this topic deserves further investigation.

\section{Massive Gravitational Solitons Via NET}
Since the implemented version of the ISM cannot provide
gravitational solitons with mass term, it is an open question
whether non--linear methods can provide such objects in 5D
low--energy effective string theories. In this Section we shall
focus on this issue. One way to approach this topic is to look for
the way of endowing the gravitational and dilaton fields with mass
and charge terms, respectively, and then, to see, for instance, if
the obtained solution corresponds to a known class. In this way it
is also possible to get new massive solutions. In order to
introduce more parameters in the constructed solitonic
configuration one can make use of a solution generating technique.
Thus, we will apply the Lie--B\"acklund transformation NET
(\ref{net}) on a seed solution which corresponds to the massless
gravitational soliton constructed in the previous Section.

The corresponding seed MEP $\X_0=\G_0+\B_0$ reads \be
\X_0=\delta^{-2}\left( \ba{ccc} -\delta^2 &0& 0\cr
4\sqrt{bc}(b+c-a\cos\theta)&-\left[r^2+b^2-(c-a\cos\theta)^2\right]
&2(b+c)r \cr 4\sqrt{bc}r &2(c-b)r &r^2+b^2-(c+a\cos\theta)^2 \ea
\right)\!\!, \label{X0}\ee where $\G_0$ and $\B_0$ are the matrix
potentials (\ref{GyB}) that parameterize the constructed solitonic
chiral matrix (\ref{N}). Here we shall use a quite simple
antisymmetric matrix $\Lambda$, namely, a matrix with just one
non--trivial constant parameter \be \Lambda=\left( \ba{ccc} 0&0&0
\cr 0&0&M\cr 0&-M & 0\ea \right). \label{l}\ee Thus, after
performing the non--linear transformation (\ref{net}) on $\X_0$ we
obtain a new solution of the theory under consideration. We shall
present the complete solution omitting all lengthy intermediate
calculations. Thus, the components of the transformed matrix
$G_{pq}$ are
\begin{eqnarray}
G_{11}&=&-\frac{\left(M^2-1\right)
\left[r^2+b^2-(c-a\cos\theta)^2\right]-4cM\left(r+Ma\cos\theta\right)}
{\left(M^2-1\right)\delta^2+4bM\left(r-Ma\cos\theta\right)},
\nonumber\\
G_{12}&=&\frac{2c\left[\left(M^2+1\right)r+2Ma\cos\theta\right]}
{\left(M^2-1\right)\delta^2+4bM\left(r-Ma\cos\theta\right)},
\\
G_{22}&=&\frac{\left(M^2-1\right)
\left[r^2+b^2-(c+a\cos\theta)^2\right]+4cM\left(r+Ma\cos\theta\right)}
{\left(M^2-1\right)\delta^2+4bM\left(r-Ma\cos\theta\right)},
\nonumber\label{G'pq}\end{eqnarray} the transformed metric
functions $\omega_{\vp}^q$ are \begin{eqnarray} \omega_{\vp}^1&=&
\frac{-2a\sqrt{bc}\left[\left(M^2-1\right)r+2M(b+c)\right]\sin^2\theta}
{\left(M^2-1\right)\left(\Delta+a^2\sin^2\theta\right)}, \nonumber\\
\omega_{\vp}^2&=&2\sqrt{bc}\left[\cos\theta-
\frac{a\left[\left(M^2+1\right)\left(b+c\right)+\left(M^2-1\right)a\cos\theta\right]\sin^2\theta}
{\left(M^2-1\right)\left(\Delta+a^2\sin^2\theta\right)}\right],
\label{oms'}\end{eqnarray} the three--dimensional dilaton field
remains the same under the NET, thus \be
e^{2\p}=1-\frac{4bc}{\Delta+a^2\sin^2\theta}. \label{3dilatonnet}\ee
The transformed components of the antisymmetric Kalb--Ramond field
are the following
\begin{eqnarray}
B_{12}&=&\frac{-2b\left[\left(M^2+1\right)r-2Ma\cos\theta\right]}
{\left(M^2-1\right)\delta^2+4bM\left(r-Ma\cos\theta\right)},
\nonumber\\
B_{4,\vp}\5&=&-2\sqrt{bc}\left\{
\frac{\left[2(b+c)M-\left(M^2-1\right)r\right]a\sin^2\theta}
{\left(M^2-1\right)\left(\Delta^2+a^2\sin^2\theta\right)}\right.
\nonumber\\
&+& \left. \frac{2b\left[\left(M^2+1\right)r-2Ma\cos\theta\right]
\left[\left(M^2\!-\!1\right)\Delta-\left(M^2+1\right)(b+c)a\sin^2\theta\right]}
{\left(M^2-1\right)\left(\Delta^2+a^2\sin^2\theta\right)
\left[\left(M^2-1\right)\delta^2+4bM\left(r-Ma\cos\theta\right)\right]}\right\},
\\
B_{5,\vp}\5&=& 2\sqrt{bc}\left\{\frac
{\left[\left(M^2\!-\!1\right)\Delta\cos\theta+\left(M^2+1\right)(b+c)a\sin^2\theta\right]}
{\left(M^2-1\right)\left(\Delta^2+a^2\sin^2\theta\right)}\right.
\nonumber\\
&+& \left. \frac
{2ab\left[\left(M^2+1\right)r-2Ma\cos\theta\right]\left[2(b+c)M+\left(M^2-1\right)r\right]\sin^2\theta}
{\left(M^2-1\right)\left(\Delta^2+a^2\sin^2\theta\right)
\left[\left(M^2-1\right)\delta^2+4bM\left(r-Ma\cos\theta\right)\right]}\right\}
\nonumber\label{B'pfi}\end{eqnarray} and the 5D dilaton field
reads \be
e^{\p\5}=\frac{\left(M^2-1\right)\left[r^2+(b-c)^2-a^2\cos^2\theta\right]}
{\left(M^2-1\right)\delta^2+4bM\left(r-Ma\cos\theta\right)}.
\label{5dilatonnet}\ee It is a straightforward exercise to check that
when $M$ vanishes, we recover the seed solitonic solution. By
studying the asymptotic behaviour of the field configuration we
can obtain information about the existence of its masses and
charges. Thus, for the components $G_{pq}$ we observe that there
exist mass terms \be G_{11}\mid_{r\rightarrow\infty}\sim
-1+\frac{2m_{11}}{r}, \quad\quad
G_{12}\mid_{r\rightarrow\infty}\sim -\frac{m_{12}}{r}, \quad\quad
G_{22}\mid_{r\rightarrow\infty}\sim 1-\frac{2m_{22}}{r},
\label{G'pqinfty}\ee where the masses $m_{pq}$ are defined as
follows \be m_{11}=\frac{2(b+c)M}{\left(M^2-1\right)}, \quad\quad
m_{12}=\frac{2c\left(M^2+1\right)}{\left(M^2-1\right)}, \quad\quad
m_{22}=\frac{2(b-c)M}{\left(M^2-1\right)}.\ee
Analogously, we see that the transformed rotation functions
$\omega_{\vp}^q$ and the three--dimensional dilaton maintain the
same asymptotic behaviour, i.e., they do not change their
behaviour at spatial infinity under the NET. The transformed
component $B_{12}$ of the antisymmetric tensor field as well as
the 5D dilaton possess Coulomb terms \be
B_{12}\mid_{r\rightarrow\infty}\sim\frac{b_{12}}{r}, \quad\quad
e^{\p\5}\mid_{r\rightarrow\infty}\sim 1+\frac{D}{r},
\ee where the new charges have been introduced
\be b_{12}=\frac{2b\left(M^2+1\right)}{\left(1-M^2\right)},
\quad\quad D=\frac{4bM}{\left(1-M^2\right)}. \label{B'fi'infty}\ee
Finally, the asymptotic behaviour of the $B_{p+3,\vp}\5$
components of the Kalb--Ramond field reads \be
B_{4,\vp}\5\mid_{r\rightarrow\infty}\,\sim
\frac{2\sqrt{bc}\left[\!-2b\left(M^2\!+\!1\right)\cos\theta\!+\!a\left(M^2\!-\!1\right)\right]
\sin^2\theta}{\left(M^2-1\right)r}, \quad \quad
B_{5,\vp}\5\mid_{r\rightarrow\infty}\,\sim 2\sqrt{bc}\cos\theta.
\label{b'pfiinfty}\ee Thus, we have obtained a stationary axially
symmetric massive gravitational field configuration coupled to
non--trivial dilaton and axion fields endowed with their
corresponding charges. Again, in order to obtain an asymptotically
flat field configuration we can set to zero either $b$ or $c$. If
$c$ vanishes we get a static inhomogeneous gravitational field
with massive components $G_{11}$ and $G_{22}$ coupled to
non--trivial  Kalb--Ramond and dilaton fields endowed with their
corresponding charges. In the case when $b$ is set to zero, we
recover a static inhomogeneous gravitational field with massive
components $G_{pq}$ and vanishing dilaton and antisymmetric
fields.

Once again we observe that when we impose the asymptotically
flatness condition, we automatically get a static field
configuration since by setting to zero the NUT--like parameter
implies the vanishing of the whole metric functions (\ref{oms'}).
Thus, if our soliton represents a rotating field configuration, it
necessarily possesses the NUT--like parameter and if we search for
an asymptotically flat solution, it necessarily becomes static.
This feature is not shared by rotating configurations in General
Relativity where, for instance, one obtains the Kerr metric from
the Kerr--NUT one when the NUT charge vanishes. This fact is a
consequence of the relationships that take place between the
constants $a$, $b$ and $c$, which are arbitrary, but not
independent each other (for instance, from (\ref{abc}) it can be
seen that $b\sim c$). It seems that this is, in turn, a
consequence of the restrictions (\ref{restrics1}) and
(\ref{restrics2}) which we imposed on the constants $p_a^k$ in
order to get a symmetric chiral matrix $\N$. It is interesting to
propose another scheme for symmetrizing the matrix $\N$ which
would avoid this strange physical behaviour of the constructed
solitonic solutions and could, in principle, provide the presence
of mass and charge terms for the fields of our configuration.

\section{Conclusions and Discussion}
In this paper we have combined two non--linear methods in order to
construct a solitonic gravitational field configuration to the 5D
low--energy bosonic sector of string theory. By following the
modified version of the ISM proposed by Yurova we clarified the
unavoidable massless character of the obtained gravitational
solitonic solutions. Therefore, by imposing suitable conditions on
the constants that parameterize the chiral matrix $\N$, we
construct a  soliton consisting of a rotating massless
gravitational field configuration coupled to a chargeless dilaton
and to an axion field endowed with charge. This solution has
similar but different physical properties and limits when compared
to the solution constructed in \cite{yurova}. Afterwards, we
provide this field configuration with mass and charge terms by
performing a simplified non--linear NET on it. Here we would like
to point out that the non--trivial physical effect of the NET is
quite different in General Relativity and string theory. It is
well known that in the framework of General Relativity, the Ehlers
transformation provides the presence of the NUT--like charge when
applied on vacuum seed solutions \cite{kramer}. However, within
the framework of string theory we observe that a simplified
version (with just one parameter) of this transformation provides
the mass and dilaton charge when applied on a massless seed
gravitational solution and does not affect the NUT parameter at
all.

Let us say few words about the physical properties of our massless
and massive gravitational field configurations. As it was
mentioned above, when looking towards a rotating black string
interpretation of them, we impose the asymptotic flatness
condition and set to zero the NUT parameter. However, when we drop
the NUT--like charge, the remaining configuration becomes static.
Thus, from one side, if our solitons are restricted to be
rotating, they necessarily possess a NUT parameter, and, from the
other side, if they are conditioned to be asymptotically flat,
they are necessarily static. This fact is due to the overall
constant factor $\sqrt{bc}$ of the functions $\omega_{\vp}^p$ for
both massless and massive solutions. Such functions define both
the angular momentum of the configurations and their NUT--like
charges. It seems that this is, in turn, a consequence of the
conditions (\ref{restrics1}) and (\ref{restrics2}) we have imposed
towards the symmetrization of the matrix $\N$. It is of interest
to propose another scheme for symmetrizing the matrix $\N$ which
would avoid this strange physical behaviour of the constructed
solitonic solutions. This peculiar physical property of our
solutions is quite strange, but on the other side, it is quite
interesting and deserves more investigation. The computation of
the scalar curvature and other invariants will help in clarifying
this point. A current investigation in this direction is also in
progress.

Within the framework of the solution generating technique using
non--linear transformations of Lie--B\"acklund type, it is
interesting to see whether the use of a full parameterized
constant matrix $\Lambda$ will affect the asymptotic behaviour of
the transformed solution and provide more independent charges. In
this way we could obtain a massive rotating solitonic object
endowed with NUT--like charge that remains spinning after the
vanishing of the NUT parameter. However, such transformation
involves really lengthy algebraic calculations which we hope to
perform in the near future. Another way of generalizing the
present results is by applying this kind of non--linear
Lie--B\"acklund transformations to string systems that include
vector fields. In this context it is the NHT which must be
performed on our massless seed solution. Thus, within the
low--energy string theory realm, this methods could, in principle,
lead to the construction of new charged black hole (black brane)
solutions in $D>4$ dimensions, where it is known that such objects
do exist \cite{othersols}. Moreover, it is interesting to apply
the ISM for the whole spectrum of the reduced to two dimensions
low--energy effective field theory of the heterotic string (taking
into account the Abelian vector fields). This could be possible
because of the mentioned above relationship between this theory
and the EM theory. For a review of the ISM applied to the
stationary axisymmetric EM system see \cite{bv}.

\section*{Acknowledgments}

One of the authors (AHA) is really grateful to the Theoretical
Physics Department of the Aristotle University of Thessaloniki
and, specially, to Prof. J.E. Paschalis for useful discussions and
for providing a stimulating atmosphere while part of this work was
being done. He also acknowledges a grant for postdoctoral studies
provided by the Greek Government. Both authors thank IMATE--UNAM
and CINVESTAV for provided library facilities while part of this
investigation was in progress. This research was supported by
grants CIC-UMSNH-4.18 and CONACYT-42064-F.


\end{document}